\documentclass[preprint2]{aastex}

\begin{document}

\title{A {\it Chandra} ACIS Observation of the X-ray Luminous SN~1988Z}

\author{Eric M. Schlegel\altaffilmark{1,2} \& Robert Petre\altaffilmark{3}}

\altaffiltext{1}{High Energy Astrophysics Division, Smithsonian Astrophysical
Observatory, Cambridge, MA 02138}

\altaffiltext{2}{Current address: Department of Physics and Astronomy,
University of Texas-San Antonio, San Antonio, TX 78249}

\altaffiltext{3}{X-ray Astrophysics Group, NASA-GSFC, Greenbelt, MD 20771}

%\altaffiltext{1}{text}

\begin{abstract}

SN~1988Z is the most luminous X-ray-emitting supernova, initially
detected in 1995 using the {\it ROSAT} HRI with a luminosity of
$\sim$8$\times$10$^{40}$ erg s$^{-1}$ \citep{FT96}. Its high
luminosity was ascribed to expansion of the blast wave into an
especially dense circumstellar medium.  In this paper, we describe a
recent observation of SN~1988Z using the ACIS detector on {\it
Chandra}.  We readily detect SN~1988Z, obtaining $\sim$30 net counts
which corresponds to a 0.2-2.0 keV luminosity of
$\sim$3.2$\times$10$^{39}$ erg s$^{-1}$.  The calculated quantiles for
the extracted counts allow a broad range of temperatures, but require
a temperature hotter than 5 keV if there is no intrinsic absorption.
The long term light curve (1995-2005) declines as t$^{-2.6\pm0.6}$.
This is one of the steepest X-ray light curves.  The X-ray luminosity
indicates that the emitting region has a high density ($>$10$^5$
cm$^{-3}$) and that the density profile is not consistent with a
constant mass loss stellar wind during the $\sim$5,000 years before
the explosion.  If the circumstellar medium is due to progenitor mass
loss, then the mass loss rate is extremely high ($\sim$10$^{-3}$
M$_{\sun}$ yr$^{-1}$(v$_w$/10 km s$^{-1}$)).  The X-ray results are
compared with the predictions of models of SN1988Z.

\end{abstract}


\section{Introduction}

The detection of X-rays from a supernova is a rare event.  While
several mechanisms might produce X-ray emission - a thermal flash
associated with breakout of the shock from the stellar surface,
Compton scattered ${\gamma}$-rays from metals synthesized in the
explosion, shocked-heated circumstellar matter - only the latter has
been detected more than once.  Down-scattered ${\gamma}$-rays have
been detected only from SN~1987A between 6 and 12 months after the
explosion.  Strong X-ray emission from shock-heated circumstellar
matter occurs only when a relatively large amount of material is
located in close proximity to the progenitor.  As this material is
generally the product of the pre-supernova stellar mass loss,
measuring the X-ray light curve reveals the mass loss history.  As the
blast wave propagates outward considerably faster than the wind, the
supernova light curve over the course of a few years reveals the mass
loss history of the progenitor's final several thousand years.

The total list of supernovae detected in X-rays consists of $\sim$24
objects, and is slowly growing as a result of {\it Chandra}, {\it
XMM-Newton} and now {\it Swift} observations.  \footnote {A complete
list of X-ray SNe and references is available at
http://lheawww.gsfc.nasa.gov/users/immler/supernovae list.html} Long
term light curves exist for about half a dozen of these \citep{IK05}.
These light curves show a range of behavior, from nearly flat (e.g.,
SN~1979C and SN~1978K since its 1991 discovery), to declining at
various rates, to more complex behavior (e.g., SN~1993J - \cite{ZA03}).
These light curves are decidedly incomplete, however, as data sets
consisting of more than four observations exist for only four SNe
(SN~1978K, SN~1993J, SN~1998S, SN~1999em) and only six supernovae have
been followed beyond a few years.  In this paper we add SN~1988Z to
this select group.

SN~1988Z is one of a small number of extremely luminous, X-ray
emitting supernovae \citep{Sch95,ImmLew03,Sch05} and is the most
distant confirmed X-ray emitter.  SN~1988Z was discovered on 1988 Dec
12 by two independent observers, C. Pollas \citep{Pollas1988} and
G. Candeo \citep{Cappellaro88}, near the galaxy MCG +03-28-022 (= Zw
095-049).  It was classified as a Type II by \cite{Cowley1988} from an
optical spectrum.  \cite{Filippenko89} confirmed the classification,
noting the spectral peculiarities of very blue color and lack of P Cyg
profiles.  \cite{Schl90} included it as a member of a new subclass,
the Type IIn supernovae.  \cite{Weiler90} reported the first radio
detection of SN~1988Z using the Very Large Array; subsequent detailed
early radio observations were reported in \cite{VanDyk93} and recent
radio observations in \cite{Williams02}. \cite{Aret99} amassed the
observations of SN~1988Z available as of late 1998 and constructed the
first spectral energy distribution for any Type IIn supernova.

SN~1988Z's importance arises from its extreme distance, $\sim$93 Mpc 
(assuming H$_o$=72 km s$^{-1}$ Mpc$^{-1}$).  \footnote{A
marginal detection of SN~2002hi, at an estimated distance of 180 Mpc,
is reported by \cite{Pooley2003}
%with 2 counts at
%the position of the supernova.  SN~2002hi is located $\sim$1$''$ from
%the host galaxy nucleus, so a portion of those counts may be emitted
%by the nucleus.  
The text is available at the Astronomer's Telegram website,
www.astronomerstelegram.org/.}.  At a discovery luminosity of
$\sim$8$\times$10$^{40}$ ergs~s$^{-1}$, it is also the most luminous.
(This luminosity was revised down to
(3$\pm$1.6)$\times$10$^{40}$~ergs~s$^{-1}$ by \cite{Aret99}).  More
importantly, SN~1988Z has been proposed by \cite{FT96} as the prototype
"compact supernova," an event in a high density medium emitting
sufficient luminosity to power the broad-line regions of some active
galactic nuclei (e. g., \cite{Terlevich1992}).  While this idea has
fallen out of favor, the existence of SN~1988Z suggests that extremely
luminous supernovae occur via the proposed mechanism, and are
interesting objects of study in their own right.  A second
interpretation suggests that the high luminosity arises from shocks
driven into dense clumps formed by an unstable stellar wind
(\cite{CD94};\cite{Williams02}).  Either mechanism requires a
substantial circumstellar envelope.

In this paper, we present the result of a {\it Chandra} ACIS 
observation of SN~1988Z. {\it Chandra's} sensitivity allows us to 
clearly detect the object a decade after its initial X-ray detection. 
While limited spectral information is available from the observation, 
the inferred flux and the 10-year light curve lead to interesting 
inferences about SN~1988Z and its environment.

Throughout this paper we adopt a distance to SN~1988Z of 93 Mpc,
calculated from the redshift of 6670 km s$^{-1}$ \citep{BCT89} and
using the value for H$_0$ of 72 km s$^{-1}$ Mpc$^{-1}$ derived from
both WMAP \citep{Spergel03} and the {\it Hubble} Space Telescope
\citep{Freedman01} .  Note that this is different from the distance
assumed in previous papers describing X-ray and radio observations
(\cite{FT96}; \cite{Aret99}; \cite{Williams02}; Table~\ref{assump_sum}).

\section{Observation}

{\it Chandra} observed SN~1988Z on 2004 June 29 (MJD 53186) for 
$\sim$50 ksec with the ACIS detector \citep{Gar03}.  The aimpoint of 
the detector fell on the back-illuminated S3 CCD.  The observation 
was cut short by $\sim$2 ksec by increasing electron contamination 
following a solar flare and during approach to the radiation belts. 
We extracted a light curve of the background, masking out all bright 
sources detectable by eye, to search for any signatures of flaring 
events known to affect back-illuminated CCDs \citep{Chan02}.  The 
background rate increased dramatically near the end of the 
observation, reflecting the increasing radiation background from the 
solar flare.  Removal of the increased background interval yielded a 
net exposure time of 44.62  ksec.

\section{Results}

\begin{figure}
\begin{center}
\caption{X-ray positions of detected sources overlaid on an blue optical
Digital Sky Survey 2 image of the field surrounding SN~1988Z.
SN~1988Z is at the center of the image; its host galaxy,
MCG +03-28-022, which is {\it not} an X-ray source, lies due west.}
%\scalebox{0.5}{\includegraphics{sn88z_field_xfig.eps}}
\scalebox{0.5}{\includegraphics{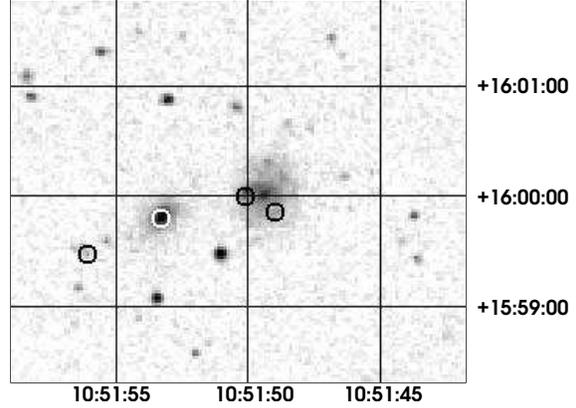}}
\label{sn88z_img}
\end{center}
\end{figure}

Figure~\ref{sn88z_img} shows the optical field containing SN~1988Z and
its host galaxy MCG~+03-28-022.  Circles indicate the X-ray sources
detected in the 0.3-10.0 keV band.  A total of four sources were
detected within 2 arc minutes of the nucleus.  The brightest of these
is located within 0$''$.5 of the radio position of SN~1988Z
\citep{Williams02}.  This source is marked in Figure~\ref{sn88z_img}
with the white circle.  It is the brightest source among the four by a
factor of two.  Note that the nucleus of the host galaxy is {\it not}
detected as an X-ray source.

We extracted counts from SN~1988Z using an aperture of 4$''$ 
centered on the source; this aperture encloses $>$98\% of the point 
spread function of an on-axis point source \citep{Chan02}.  We also 
extracted counts in several bands; for example, to match the bandpass 
of the {\it ROSAT} HRI \citep{Aret99}, we used the 0.2-2 keV band. 
There is a larger uncertainty (and fewer net counts) in the 0.2-2 keV 
band than in the 0.3-2 keV band because of the higher background 
below 0.3 keV \citep{Chan02}.

The background was extracted from a circle of radius 20$''$ centered
$\sim$30$''$ to the southeast of the source circle to avoid any
possible emission from the host galaxy.  The net total-band count rate
is (8.4$\pm$1.7)$\times$10$^{-4}$ counts s$^{-1}$.  The rates for
all extracted bands are listed in Table~\ref{counts}.  At these low
count rates, the effects due to event pileup are negligible
\citep{Davis01}.

\subsection{Spectral Constraints}

Quantiles have been demonstrated to be more robust than hardness
ratios for estimating spectral properties of sources with very limited
numbers of counts because they do not generate upper limits
\citep{Hong04}.  For the ACIS observation, we extracted the photon
energy values for the events in the source and background apertures
and generated the 25\% and 75\% quantile values.

\begin{figure}
\begin{center}
\caption{Quantile plot for SN~1988Z using the description of
\cite{Hong04}.  The underlying grid represents a MeKaL model of a
coronal equilibrium plasma; the 'mountainous' shape arises from
changing line emission as (primarily) the temperature parameter is
varied.  The N$_{\rm H}$ axis starts at 0.001$\times$10$^{22}$
cm$^{-2}$ with isolines at 0.01, 0.1, 0.5, 1.0, and
5.0$\times$10$^{22}$ cm$^{-2}$.  The temperature axis shows isothermal
lines of 0.2, 0.4, 1, 2, 5, and 10 keV.  The position of SN~1988Z
corresponds approximately to N$_{\rm H}$ $\sim$0.5$\times$10$^{22}$
cm$^{-2}$ and kT $\sim$5 keV.  The temperature range includes
$\sim$3-10 keV; the N$_{\rm H}$ range is $\sim$0.01-2$\times$10$^{22}$
cm$^{-2}$.}
%\scalebox{0.35}{\rotatebox{-90}{\includegraphics{sn88z_quant.ps}}}
\scalebox{0.35}{\rotatebox{-90}{\includegraphics{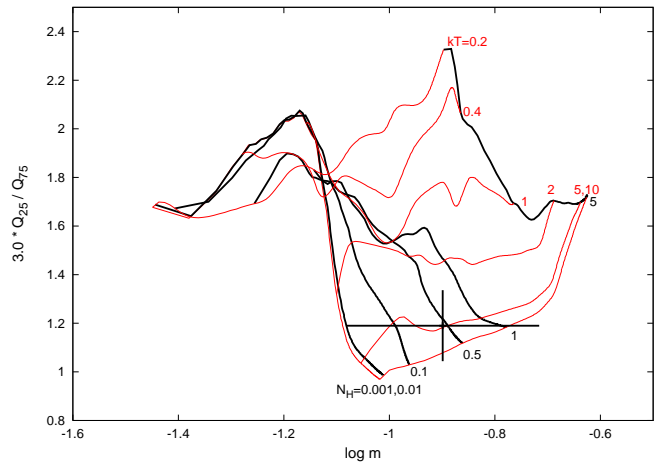}}}
\label{quantpic}
\end{center}
\end{figure}

Figure~\ref{quantpic} shows the result of quantile analysis for a
coronal ionization equilibrium plasma (MeKaL) grid.  The MeKal grid
has the depicted shape because of the changing presence of line
emission as the temperature (primarily) is varied.
%For the bremsstrahlung model, the quantile analysis yields a 
%preferred temperature of
%approximately 2.5 keV with a 1-sigma range of 0.8-6 keV.  The best 
%column density estimate
%is 1.3$\times$10$^{21}$ cm$^{-2}$, with a 1-sigma range between
%3$\times$10$^{20}$ cm$^{-2}$ and 1$\times$10$^{22}$ cm$^{-2}$.
%The allowed temperature range includes the1 keV values assumed by
%\cite{FT96} and \cite{Aret99}.
%For the MeKaL model, the
The quantile analysis yields a preferred temperature of $\sim$2-3~keV, but 
the 90\% uncertainty covers a very broad range, leaving the 
temperature high, but essentially undefined.  The preferred column density is 
5$\times$10$^{21}$ cm$^{-2}$, with a range between 1$\times$10$^{19}$ 
cm$^{-2}$ and 2$\times$10$^{22}$ cm$^{-2}$.  The broad allowed 
temperature range can be understood from the relative instability of 
quantiles using the MeKaL model caused by the emergence and decline of various 
lines as the temperature varies.  If we externally constrain the 
value of the column density by using the Galactic value of 
1.4$\times$10$^{20}$ cm$^{-2}$, then the temperature is constrained 
to be hotter than $\sim$5~keV.  

\subsection{Flux and Light Curve}

Converting the measured count rates into physical units requires estimates 
of distance and column density plus an assumption about the emission 
mechanism.  We use different values for all three from those used in 
previous papers (\cite{FT96}, \cite{Aret99}), as justified below.

We estimated the column density toward SN~1988Z using the dust maps of
\cite{SFD98} and the E$_{\rm B-V}$-N$_{\rm H}$ relation of
\cite{PS95}.  Together these yield a value of 1.4$\times$10$^{20}$
cm$^{-2}$, lower than the 2.5$\times$10$^{20}$ cm$^{-2}$ from the
lower angular resolution maps of \cite{BH82} used by \cite{Aret99}.
\cite{Aret99} point out that the X-ray column density in the immediate
vicinity of the supernova could be much higher if the pre-ionized
material enshrouding the supenova is only partially ionized, as the
partially ionized material will absorb a substantial fraction of the
soft X-ray flux.  They estimate that the column density could easily
be as high as 10$^{22}$ cm$^{-2}$.  The absence of absorption in the
radio data suggest this is not the case \citep{Williams02}.
Nevertheless it should be noted that the soft X-ray luminosity would
be substantially underestimated if we adopt the Galactic column
density while the supernova instead has a high intrinsic column.

A supernova shock encountering dense circumstellar material or ejecta
generally produces a line-rich X-ray spectrum from atomic transitions
within nearly stripped ions.  Given the inferred density of the
shocked medium around SN~1988Z, it is expected to rapidly attain
collisional ionization equilibrium \citep{CF94,ChevFran04}.  We
therefore use the MeKaL model for estimating flux and luminosity
\citep{Liedahl92, Mewe85}.  This is also different from previous
publications, which used thermal Bremsstrahlung
\citep{FT96,Aret99}(Table~\ref{assump_sum}).  If SN~1988Z emits
copious hard X-rays, then it is possible that the entire surrounding
medium has become fully ionized, in which case Bremsstrahlung emission
would dominate.  Given the absence of spectral information, either
spectral model can be used equally well to infer relative fluxes among
the available X-ray observations.

The quantile analysis indicates that if we externally constrain the
value of the column density by using the Galactic value of
1.4$\times$10$^{20}$ cm$^{-2}$, then the temperature is constrained to
be hotter than $\sim$5~keV.  To maintain consistency with the quantile
results we assume for our flux estimates a 5~keV plasma with Galactic
column density.  Flux analysis of the X-ray emission from other
remnants with higher flux has generally used a temperature of 0.8 keV,
inferred from spectral fitting \citep{Sch05, Imm05}.  This has been
intepreted as the temperature of the reverse shock.  It should be
noted that the emissivity of a plasma in the 0.2-2.0 keV band varies
slowly over the temperature range from 0.5 to 10 keV.  The emissivity
for a 0.8 keV plasma is approximately 25 percent higher than a 5 keV
plasma.  The acknowledged inconsistency with other analyses resulting
from the use here of the higher temperature is however comparable with
or smaller than other uncertainties, such as possible temperature
evolution between the {\it ROSAT} and {\it Chandra} observations.

The recession velocity of MCG~+03-28-022 is 6670 km s$^{-1}$
\citep{BCT89}.  Adopting the value for H$_0$ of 72 km s$^{-1}$
Mpc$^{-1}$ consistent with with {\it WMAP} and {\it Hubble} Space
Telescope measurements \citep{Spergel03, Freedman01} yields a distance
to SN~1988Z of 93 Mpc.  Previous papers (\cite{FT96}, \cite{Aret99},
Table~\ref{assump_sum}) assumed a smaller Hubble Constant value (H$_0$
= 50 km s$^{-1}$ Mpc$^{-1}$) and thus inferred a larger distance (and
thus a higher luminosity).

Table~\ref{fluxes} lists the inferred 0.2-2.0 and 0.5-2.0 keV fluxes
and luminosities for the {\it ROSAT} and {\it Chandra} observations.
The inferred flux for the {\it ROSAT} observations is essentially
unchanged, as the reduction due to use of the lower Galactic N$_{\rm
H}$ is offset by the increase due to the higher kT.  The luminosities
for the {\it ROSAT} observations are reduced from those in
\cite{Aret99} by about a factor of two as a consequence of the revised
distance.  From the original estimate of
2$\pm$0.7$\times$10$^{41}$~ergs~s$^{-1}$ \citep{FT96}, and the
adjusted value of 8$\pm$3.2$\times$10$^{40}$~ergs~s$^{-1}$
\citep{Aret99}, the peak observed 0.2-2.0~keV luminosity is now
3.4$\pm$1.2$\times$10$^{40}$~ergs~s$^{-1}$.  Of course, estimates of
luminosity in other bands also drop by a factor of two as a result of
the revised distance. Despite this downward revision, SN~1988Z remains
the most luminous supernova with a confirmed X-ray detection.

\begin{figure}
\begin{center}
\caption{X-ray light curve in the 0.2-2  keV band based upon
  {\it ROSAT} HRI and {\it Chandra} ACIS observations.  The 
data 
points have all been converted to flux using
the same model.  The best fit power law yields a light curve 
f$\propto$t$^{-2.6}$.  
Also shown is a power law slope of -11/7, as 
predicted by \cite{Aret99}.}
%\scalebox{0.35}{\rotatebox{-90}{\includegraphics{sn88z_log_lc_final.ps}}}
\scalebox{0.35}{\rotatebox{-90}{\includegraphics{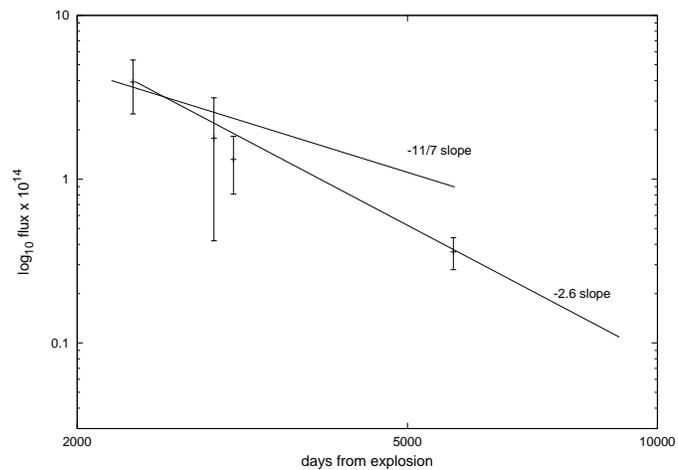}}}
\label{lc88z}
\end{center}
\end{figure}

The 0.2-2.0 keV fluxes are displayed graphically in 
Figure~\ref{lc88z}, the X-ray light curve. The ACIS flux is an order 
of magnitude lower than the initial {\it ROSAT} HRI flux.  A fit to 
dlog(flux)/dlog(t) for the 0.2-2.0 keV band yields a slope of 
-2.6$\pm$0.6.  Thus the luminosity evolution of SN~1998Z in this band 
follows the curve L$_{\rm X}$$\propto$~t$^{-2.6\pm0.6}$.
SN1970G, with a slope of 2.7$\pm$0.9 is the only X-ray supernova with
a light curve as steep, but much later in its evolution \citep{IK05}.

\subsection{Inferred X-ray Properties}

It is possible to infer information about the density of the gas from
the X-ray luminosity, if an assumption is made about the size of the
X-ray emitting region.  \cite{Aret99} show that the H$\alpha$ width
decrease is consistent with the shock velocity being proportional to
t$^{-5/7}$.  If we assume that the H$\alpha$ width is a measure of the
shock velocity, it is possible to estimate the radius and hence the
volume of the region enclosed by the shock for each the four X-ray
observations.  The inferred velocity, radius and volume are listed in
Table~\ref{density}.

For a thermal plasma, L$_x$~=~$\Lambda$(T)n$_e^2$V$_x$, where P(T) is
the plasma emissivity, n$_e$ the electron density, and V$_x$ the
volume of the emitting material.  The product n$_e^2$ V$_x$ is
referred to as the volume emission measure.  For a plasma with a
temperature in the range 0.5-10.0 keV, the emissivity in the 0.2-2.0
keV band is $\sim$10$^{23}$ erg~cm$^3$~s$^{-1}$ \citep{RCS76}.  In
Table~\ref{density} we have tabulated the value of n$_e^2$ V$_x$ for
each of the observations.  The emitting volume V$_x$ is equal to
V$\phi$, where V is the total volume enclosed by the shock and $\phi$
is the volume filling factor.  Table~\ref{density} indicates that the
value of n$_e$$\phi^{1/2}$, the density of the shock-compressed gas,
varies from 6.2$\times$10$^5$ cm$^{-3}$ for the first {\it ROSAT}
observation to 1.1$\times$10$^5$ cm$^{-3}$ for the {\it Chandra}
observation.  Thus the density is high, even for the lowest density
case, assuming the plasma uniformly fills the shocked volume.  Prior
to being shock heated, the density of the ambient gas will be at least
a factor of 4 lower than the post shock density.

In Table~\ref{density} we also list the inferred X-ray emitting mass.
In all cases, the minimum mass is several solar masses.  The high
shock velocity ($>$1,000~km~s$^{-1}$) ensures that the bulk of the
luminosity emerges in the X-ray band throughout the interval covered
by the {\it ROSAT} and {\it Chandra} observations.

The methodology used above for inferring properties of the X-ray
emitting gas are similar to those first used by Immler and co-workers
(as summarized in \cite{ImmLew03}).  The fundamental difference is
that these authors assume a constant shock velocity, whereas we assume
one decreasing with time.  If we were to assume a constant velocity,
the values we infer for density and mass would be similar, but would
be higher or lower, depending upon the velocity we assume.

The rapid decrease of the X-ray emitting mass can in principle be used
to set a lower limit on the density and thus constrain the filling
factor.  A decline in the emitting mass indicates that the
shock-heated material is cooling quickly.  The cooling time t$_c$ can
be written as: t$_c$~=~0.2 v$_8^2$/(n$_7$$\Lambda_{23}$)~yr, where
v$_8$ is the shock velocity in units of 10$^8$ cm s$^{-1}$, n$_7$ is
the density in units of 10$^7$ cm$^{-3}$, and $\Lambda_{23}$ is the
plasma emissivity in units of 10$^{-23}$ erg~cm$^{3}$~ s$^{-1}$
(\cite{Aret99}).  The contents of Table~\ref{density} indicate
n$_7$$\ge$0.01 and v$_8$$\ge$1, adopting $\Lambda_{23}\sim$1.  That
the X-ray emitting mass is decreasing at a rate between $\sim$0.2 and
$\sim$1 solar mass per year suggests a cooling time on the order of
1-5 years.  Such a cooling time requires a density $>$4$\times$10$^5$~
cm$^{3}$.  As this density lower limit is consistent with the inferred
X-ray emitting gas density, we conclude that a filling factor as large
as unity is allowed.

It is also possible to estimate the mass loss rate of the progenitor,
if we assume that the circumstellar envelope is due to mass loss.
Integrating the X-ray emitting mass yields a variable mass loss rate.
For the epoch of the initial {\it ROSAT} observation, corresponding to
$\sim$4000$\times$(v$_w$/10 km s$^{-1}$) years before the explosion,
the minumum mass loss rate (from assuming $\phi$=1) is
$\sim$1.2$\times$10$^{-3}$ M$_{\sun}$ yr$^{-1}$v$_w$/10 km s$^{-1}$).  For
the epoch of the {\it Chandra} observation, which corresponds to
$\sim$5500$\times$(v$_w$/10 km s$^{-1}$) years before the explosion,
the minumum mass loss rate is $\sim$4$\times$10$^{-4}$ M$_{\sun}$
yr$^{-1}$(v$_w$/10 km s$^{-1}$).  Alternatively, the X-ray cooling
rate requires at least 7 M$\sun$ to have been swept up.  This
corresponds to an average mass loss rate of $\sim$1.3$\times$10$^{-3}$
M$_{\sun}$ yr$^{-1}$(v$_w$/10 km s$^{-1}$).  This range of mass loss
rates lie at the extreme end of those inferred for X-ray supernovae
\citep{ImmLew03}, as would be expected given its high luminosity.

We assumed above that the gas is in collisional ionization
equilibrium.  For a shock-heated plasma, CIE is attained when the
product n$_e$t$>$10$^{13}$cm$^{-3}$ s.  The inferred gas densities
require t$\sim$1-8$\times$10$^8$$\phi^{1/2}$ s = 3-24$\phi^{1/2}$ yr
to reach equilibrium.  Thus the equilibration time is comparable with
the cooling time, and the assumption of CIE is justified.

The X-ray fluxes require n$_e$$\phi^{1/2}$ to decline with time.
Since L$_x$$\propto$ t$^{-2.6}$ and n$_e$$\phi^{1/2}$$\propto$
L$_x^{-1/2}$, we expect n$_e$$\phi^{1/2}$$\propto$ t$^{-1.3}$.  The
data do not allow determination of whether the density declines while
the filling factor remains constant, the filling factor declines while
the density remains constant, or both change with time.  In the
discussion below, we consider the implications of a decline of density
and filling factor in the context of models proposed for SN~1988Z.

\section{Discussion}

Two distinct models have been proposed to explain the multiband light 
curve of SN 1988Z.  One postulates that the supernova exploded in an 
extremely dense ($\sim$10$^7$~cm$^{-3}$) medium (\cite{FT96}; 
\cite{Aret99}).  As the supernova shock propagates through such a 
medium, it loses energy rapidly, and evolves in a matter of years to 
the "shell forming," or radiative stage.  Extremely high luminosities 
are produced as the SN radiates away most of its kinetic energy over 
decades.  The second model postulates propagation of the SN shock 
through a clumpy medium, formed by an unstable wind from a progenitor 
star that underwent extreme mass loss (\cite{CD94}; 
\cite{Williams02}).  In this case the forward shock propagates 
rapidly outward, driving slower shocks into the clumps it encounters. 
X-ray (and radio) emission arises from the shocked clumps.  Below we 
address how the predictions of these two models compare with the 
{\it Chandra} observation and the inferred X-ray light curve.  

%It should be 
%emphasized that in either case SN~1988Z is unique - none of the other 
%23 known X-ray emitting SNe shows evidence for either an extremely 
%dense or a clumpy surrounding medium \citep{IK05}.

\cite{Aret99} provide a capsule summary of the so-called "compact 
supernova" model, and compare optical, radio, and X-ray data from 
SN~19988Z with the model predictions.  They show that the light curves 
in many bands (for the first decade after the explosion) follow 
trends similar to the model predictions.  They also show that the 
shock velocity, which they infer from the width of the H$\alpha$ 
line, follows the t$^{-5/7}$ behavior predicted from the model.  (We 
have used this inference above in our estimate of the radius enclosed 
by the shock.)  They predict that the bolometric luminosity should 
decline as t$^{-11/7}$.  As their model also predicts that the bulk of 
the luminosity is emitted in the X-ray band, the X-ray luminosity 
should follow this decay curve.  The {\it ROSAT} fluxes appear to fit this 
pattern.  They estimate an average density on the order of 
10$^7$~cm$^{-3}$.

This model runs into difficulty when it is compared with the extended
X-ray light curve.  The measured t$^{-2.6\pm0.6}$ is not consistent
with the prediction.  The assumption of a constant density of
$\sim$10$^7$~cm$^{-3}$ is challenged by the decline in n$\phi^{1/2}$
required by the X-ray flux measurements.  A constant density of
$\sim$10$^7$~cm$^{-3}$ requires a small filling factor that decreases
with time (from 3.6$\times$10$^{-3}$ at the time of the first {\it
ROSAT} observation to 1.2$\times$10$^{-4}$ at the time of the {\it
Chandra} observation).  If the X-ray emitting gas is confined to a
shell, its thickness also decreases, from $\sim$2$\times$10$^{15}$~cm
to $\sim$1$\times$10$^{14}$~cm.  Additionally, the model is
inconsistent with the late-time radio spectrum, which shows little
free-free absorption \citep{Williams02} from the intervening material
composing the cocoon in which the supernova exploded.

There are two ways to potentially reconcile these difficulties.
First, it is conceivable that the bulk of the luminosity has shifted
out of the X-ray band.  This is not likely, based on the inferred
shock velocity from the H$\alpha$ line width measurements.  Even if
the velocity is overestimated by a factor of two at the time of the
{\it Chandra} measurement, the bulk of the emission should still arise
in the X-ray band.  Second, it is possible that the supernova shock
has encountered a change in the average medium density.  This
possibility cannot be ruled out, given the small number of
measurements postdating \cite{Aret99}.  For instance, it is possible
that the shock recently broke through the dense cocoon into a
low-density medium.  In that case, the X-ray light curve, which until
the {\it Chandra} observation could be characterized by a -11/7 slope,
might show a pronounced break.  If the change occurred at day 4500,
roughly midway between the final {\it ROSAT} and the {\it Chandra}
observations, then the light curve from day 4500 to the {\it Chandra}
observation would have and extremely steep decline: L$_{\rm
X}$$\propto$~t$^{-7}$.  The two steepest light curves observed are
those of this remnant and SN~1970G \citep{IK05}.  Therefore, while the
{\it Chandra} flux measurement does not exclude the compact supernova
model, the required steep late-time decline entails significant
modification of the model.

\cite{Williams02} champion the alternative model, whereby the high
radio luminosity is produced by slow shocks driven into dense clumps
of wind material.  They assume a constant shock velocity through the
interclump medium (20,000~km~s$^{-1}$, as was observed early in the
evolution of SN~1988Z --\cite{Turatto93}). Using this assumption, and
the theoretical framework of \cite{CD94}, they interpret the radio
light curve as arising from a medium with a low cloud number density
(filling factor).  They quantify neither the filling factor, nor the
density of the clumps or interclump medium, but they infer a mass loss
rate of 1.2$\times$10$^{-4}$ M$_0$~yr$^{-1}$, and a corresponding
${\dot M}$/w$_{wind}$ = 7.3$\times$10$^{13}$ ~g~cm$^{-1}$.

The key observation driving this interpretation is that the radio
light curve between days 1750 and 5000 has a -2.7 slope.  This slope
is essentially identical to the X-ray decline between days 2355 and
5678.  While SN radio and X-ray light curves are generally correlated,
and all X-ray supernovae are detected in the radio,
%\footnote {The
%single possible exception is SN~1999gi, an unconfirmed, marginal
%detection \citep{S01} }), 
none has been observed to have exactly the same slope (\cite{IK05}).
In fact, the X-ray light curve should have a shallower slope.  Since
the X-ray and H$\alpha$ fluxes depend on n$_e^2$ while the radio light
curve depends on n$_e^{1.4}$, the X-ray light curve should follow the
H$\alpha$ light curve, which \cite{Williams02} show is consistent with
a -2.1 slope.  For our case, since the filling factor decreases while
the volume increases with time, a simple interpretation might be a
shock running through a clumpy medium.  Based on the clumpy medium
model, a light curve with slope -2.1 is marginally consistent with the
X-ray data.

As a matter of interest, \cite{Williams02} show that the radio and
H$\alpha$ light curves steepened significantly after day 1750.  They
interpret this steepening as evidence for a decrease in the number
density of clumps.  As none of the X-ray observations were performed
before day 1750, we do not know whether the X-ray light curve also
changed slope.  Whatever the physical reason behind it might be, the
fact that one slope change has been observed, however, supports the
possibility of a second steepening during the last few years.  All the
observations listed in \cite{Williams02} were performed before day
4500.  If we assume that the X-ray light curve followed a -2.1 slope
until that time (allowed by the {\it ROSAT} fluxes), and then
decreased more rapidly after that date, then a slope of at least -2.8
is required for consistency with the {\it Chandra} flux point, which
corresponds to a radio slope of -3.9.  Again, no X-ray light curve 
this steep has ever been measured.

In summary, neither model is consistent with all of the observations.
Aretxaga et al. (1999) assume a dense, uniform shell.  This should
result in self-absorption of the radio emission which becomes smaller
with time. Williams et al. (2002) show that the self-absorption is
small, and constant, suggesting radio emission is not being produced
within a dense cocoon. If the intense UV and X-ray radiation keep this
material ionized, then little or no intrinsic absorption of soft
X-rays might be detected at this time.  The highest allowed column
density from the quantile analysis is 5$\times$10$^{22}$ of neutral
material.  \cite {Williams02}, on the other hand, assume a constant,
high shock velocity of 20,000 km s$^{-1}$.  This is not consistent
with the smooth velocity decline suggested by the H$\alpha$ width.
From the X-ray standpoint, the light curve is at odds with both
models.  Both predict flatter X-ray decays than observed: -11/7, and
-2.1 compared with the measured -2.6.  But the current X-ray
observations do not rule out either model.

The ambiguity presented by the limited X-ray data could be largely 
removed by a single additional observation.  Such an observation will 
show whether the light curve has a shallower average slope than 
inferred from the existing X-ray observations, continues to track the 
-2.6 slope, or steepened after the {\it ROSAT} observations.  If the slope 
is flatter, then it could support either model and possibly provide a 
discriminator between them, depending on whether it is closer to the 
-11/7 predicted by the compact supernova model or -2.1 predicted by 
the clump model.  As described above, a change in slope could rescue either 
model, but would require a change in the properties of the medium 
being encountered by the shock.  If the slope is constant at -2.6, 
then a different model must be sought which explains the X-ray light 
curve in the context of the numerous other observations. 
Additionally, radio observations are needed to show whether the X-ray 
and radio light curves continue to follow each other.

It is possible to compare the inferred density profile with those of
other X-ray emitting SNe.  \cite{Imm05} plot the density profiles for
several X-ray SNe, and show\footnote{Note that in the cited paper, the
density is represented by ${\rho}$; we made the change to conform to
our notation.} that $n_{CSM}\propto$~r$^{-s}$, where {\it s} ranges
between 1 and 2.  The inferred density (or more precisely
$n{\times}{\phi}^{1/2}$) follows a r$^{-1.3}$ profile, entirely within
this range.  In calculating their density profiles, \cite{Imm05}
assume a constant shock velocity.  In contrast to this assumption, we
have assumed that the shock velocity profile of SN~1988Z is known, and
that the shock decelerates. A comparison with the inferred density
for, e.g., SN~1979C indicates that its inferred density range is
comparable to the range of $n{\times}{\phi}^{1/2}$ for SN~1988Z.  This
could be interpreted as suggesting that the actual density is
substantially higher, and thus $\phi$ is much less than unity (or that
the shock velocity decreases in the other SNe, leading to different
density profiles).

\section{Conclusions}

An ACIS observation yields a clear detection of SN~1988Z 16 years
after discovery, and a decade after the first X-ray detection.
Although limited spectral information is available, the X-ray light
curve constructed using this observation and the three archival {\it
ROSAT} observations yields a flux decline proportional to
t$^{-2.6\pm0.6}$.  This decline is steeper than predicted by either of
the two prominent competing models.  The inferred density is high,
larger than 10$^5$~cm$^{-3}$, indicating a massive envelope around the
progenitor, and requiring rapid evolution of the X-ray emitting
plasma.  If the circumstellar envelope is the result of presupernova
mass loss, a very high mass loss rate is required, as might be
expected from this, the most X-ray luminous supernova.  An additional
observation can potentially discriminate between the two competing
models, or require a new model.

\acknowledgements

The authors would like to thank Stefan Immler for his helpful comments
on the manuscript.  The research of EMS was supported by contract
number NAS8-39073 to SAO.

\begin{table*}
\begin{center}
\caption{Summary of Distance, Column Density, and Model Assumptions}
\label{assump_sum}
\begin{tabular}{lrrrrr}
          &                &  D  & N$_{\rm H}$ &     & Inferred L$_{\rm X}$\tablenotemark{a} \\
Authors   &  H$_0$         & Mpc & cm$^{-2}$ & Model &  erg s$^{-1}$\\  \hline
Fabian \& Terlevich (1996) & 50 & 131 & 3$\times$10$^{20}$ & 1 keV bremss & 1$\times$10$^{41}$ \\
                           &    &     &                    & 5 keV bremss & 2$\times$10$^{41}$ \\
Aretxaga et al. (1999)     & 50 & 131 & 2.5$\times$10$^{20}$ & 1 keV bremss & 8$\times$10$^{40}$ \\
This paper                 & 72 & 93  & 1.4$\times$10$^{20}$ & 5 keV MeKaL & 3.4$\times$10$^{40}$ \\ \hline
\end{tabular}
\end{center}
\tablenotetext{a}{For the bandpass of 0.2-2 keV.}
\end{table*}

\begin{table}
\begin{center}
\caption{ACIS Count Rates for SN~1988Z in various bands}
\label{counts}
\begin{tabular}{ccc}
        &   Net   &           \cr
  Band  & Counts  &  Rate\tablenotemark{a} \cr \hline
  0.2-2.0 & 29.8$\pm$6.82 & 6.68$\pm$1.53  \cr
  0.3-2.0 & 30.4$\pm$6.82 & 6.81$\pm$1.53  \cr
  0.3-10.0 & 37.4$\pm$7.71 & 8.38$\pm$1.73 \cr
  0.5.2.0 & 29.6$\pm$6.60  & 6.64$\pm$1.49  \cr
  2.0-10.0 & 7.7$\pm$5.13 & 1.73$\pm$1.15  \cr \hline
\end{tabular}
\tablenotetext{a}{Units = 10$^{-4}$ counts s$^{-1}$.}
\end{center}
\end{table}

\begin{table*}
\begin{center}
\caption{0.2-2 and 0.5-2 keV Fluxes for SN~1988Z}
\label{fluxes}
\begin{tabular}{lllllllll}
           &  &  & &  &  Count & 
\multicolumn{2}{c}{Flux\tablenotemark{d}} & 
{Luminosity\tablenotemark{e}}\cr
  Detector & Obs. Date\tablenotemark{a}  & MJD   & Age\tablenotemark{b}& ExpT & Rate\tablenotemark{c} & 0.2-2.0 & 0.5-2.0 & 0.2-2.0 \cr \hline
  HRI  & 1995 May 16-25\tablenotemark{a} & 49884 & 2335 & 12287  & 11$\pm$4      & 3.92 & 2.75  & 3.4\cr
  HRI  & 1996 Dec 14\tablenotemark{a}    & 50725 & 2924 & 6739   & ~5$\pm$4      & 1.78 & 1.25 & 1.5 \cr
  HRI  & 1997 May 13-24\tablenotemark{a} & 50908 & 3085 & 34300  & ~3.7$\pm$2    & 1.32 & 0.92 & 1.15\cr
  ACIS & 2004 Jun 29                     & 50908 & 5678 & 44618  & ~6.7$\pm$1.5  & 0.36 & 0.25 & 0.27\cr \hline
\end{tabular}
\tablenotetext{a}{MJD at center of observation when spanning multiple
days; those observations are footnoted.}
\tablenotetext{b}{units = days; based upon adopted date of maximum =
1988 Dec 12 = MJD 47508.}
\tablenotetext{c}{Units are 10$^{-4}$ sec$^{-1}$ for 0.2-2.4 keV band.}
\tablenotetext{d}{All fluxes in 0.2-2.0 or 0.5-2 keV band with units
of 10$^{-14}$ erg s$^{-1}$ cm$^{-2}$.  Adopted model is an absorbed
vmekal, kT = 5.0 keV, N$_{\rm H}$ = 1.4$\times$10$^{20}$ cm$^{-2}$.
For the HRI values, fluxes were computed from the model after
adjusting the model normalization to match the observed count rate.}
\tablenotetext{e} {Luminosity units are 10$^{40}$ erg s$^{-1}$ for 
0.2-2.0 keV band.}

\end{center}
\end{table*}

\begin{table*}
\begin{center}
\caption{Inferred Emission Properties of SN~1988Z}
\label{density}
\begin{tabular}{lllllll}
  {Age\tablenotemark{a}} & {Velocity\tablenotemark{b}} & 
{Radius\tablenotemark{b}} & {n$^2{\phi}$V\tablenotemark{c}} & {n${\phi}^{1/2}$} & 
{$\phi_7$\tablenotemark{d}} & M$_x$ \cr 
(days) & (km s$^{-1}$) & 
(pc) & (10$^{63}$ cm$^{-3}$) & (10$^5$ cm$^{-3}$) & & M$_{sun}$ \cr \hline
2335 & 1900 & 0.044 & 3.4   & 5.7 & 3.2$\times$10$^{-3}$ & 4.9\cr
2924 & 1620 & 0.046 & 1.4   & 3.4 & 1.2$\times$10$^{-3}$ & 3.7\cr
3085 & 1560 & 0.047 & 1.15  & 3.0 & 9.1$\times$10$^{-4}$ & 3.2\cr
5678 & 1010 & 0.056 & 0.27  & 1.1 & 1.2$\times$10$^{-4}$ & 2.1\cr \hline
\end{tabular}
\tablenotetext{a}{MJD at center of observation when spanning multiple
days; those observations are footnoted.}
\tablenotetext{b}{Estimated using velocity profile in \cite{Aret99}.}
\tablenotetext{c}{$\phi$ is the dimensionless volume filling factor.}
\tablenotetext{d}{Volume filling factor assuming n=10$^7$ cm$^{-3}$
\citep{Aret99}.  The volume filling factor changes by a factor of 100
for a change in the assumed number density by a factor of 10.}
\end{center}
\end{table*}
\section{Figures}

 \end{document}